\documentclass[%
reprint,
 amsmath,amssymb,
 aps,
 prb,
 floatfix
]{revtex4-2}

\usepackage[english]{babel}
\usepackage{float}
\usepackage{comment}
\usepackage[dvipsnames]{xcolor}
\usepackage{graphicx}
\usepackage{dcolumn}
\usepackage{bm}
\usepackage{hyperref}

\begin{document}


\title{Moir\'{e} phonons in graphene/hexagonal boron nitride moir\'e superlattice}

\author{Lukas P. A. Krisna}
\affiliation{Department of Physics, Osaka University,  Osaka 560-0043, Japan}
\author{Mikito Koshino}
\affiliation{Department of Physics, Osaka University,  Osaka 560-0043, Japan}
\date{\today}


\begin{abstract}
We theoretically study in-plane acoustic phonons of graphene/hexagonal boron nitride  moir\'{e} superlattice by using a continuum model. 
We demonstrate that the original phonon bands of individual layers are strongly hybridized and reconstructed into moir\'e phonon bands consisting of dispersive bands and flat bands.
The phonon band structure can be effectively described by a spring-mass network model to simulate the motion of moir\'e domain walls, where the flat-band modes are interpreted as vibrations of independent, decoupled strings.
We also show that the moir\'e phonon has 
angular momentum due to the inversion symmetry breaking by hBN, with high amplitudes concentrated near narrow gap region.
Finally, we apply the same approach to twisted bilayer graphene,
and we find a notable difference between the origins of the flat-band modes in G/hBN and TBG, reflecting distinct geometric structures of domain pattern.
\end{abstract}


\maketitle

\section{\label{sec:1-intro}Introduction}

Moir\'e pattern plays an essential role in the physical properties
of van der Waals multilayer systems.
In twisted bilayer graphene (TBG), the electronic properties strongly depend on the twist angle \cite{LopesdosSantos2007Graphene,Li2009Observation,Laissardiere2010Localization,Shallcross2010Electronic,SuarezMorell2010Flat,Luican2011Single,Bistritzer2011Moire,Moon2012Energy,LopesdosSantos2012Continuum}, where nearly-flat bands with associated exotic correlated phenomena emerge at a magic angle ($\sim1^\circ$)  \cite{Cao2018Unconventional,Cao2018Correlated,Polshyn2019Large,Cao2020Strange,Jaoui2022Quantum}. 
At the same time, phonons in TBG are also significantly affected by the moir\'e superlattice modulation \cite{CamposDelgado2013Raman,Koshino2019Moire,Ochoa2019Moire,Lamparski2020Soliton,Gadelha2021Localization,Ochoa2022Degradation,Lu2022Low}.
At low-frequency, particularly, it was predicted that the in-plane acoustic phonons are reconstructed into moir\'{e} phonons corresponding to effective oscillations of the moir\'e pattern, where opening of moir\'e gaps and flattening of some specific bands take place \cite{Koshino2019Moire}.
Such modifications of phonon bands are expected to strongly affect the electronic \cite{Wu2018Theory,Lian2019Twisted,Wu2019Phonon,Ishizuka2021Purcell,DasSarma2022Strange} and thermal \cite{Han2021Twist,Qian2021Phonon} transport properties. 

A wide variety of 2D materials offers a playground to explore different types of moir\'{e} phonons.
For twisted transition-metal dichalcogenide bilayers, 
the moir\'{e}-induced phonon renormalization effect was studied \cite{Huang2016Low,Lin2018Moire,Maity2020Phonons,Quan2021Phonon,Lu2022Low},
and a chiral nature of phonons due to broken inversion symmetry was predicted \cite{Suri2021Chiral,Maity2022Chiral}.
The study of moir\'{e} phonons has also been extended to twisted bilayer hexagonal boron nitiride \cite{Moore2021Nanoscale}, twisted trilayer graphene \cite{Samajdar2022Moire,Gao2022Symmetry} and twisted multilayer graphenes \cite{Gao2022Symmetry,Lin2022Intralayer}.

In this paper, we investigate low-energy moir\'{e} phonons in graphene on hexagonal boron nitride (G/hBN), as the first example of hetero bilayer systems. While bulk hBN is commonly used as a substrate for two-dimensional materials to achieve high mobility \cite{Dean2010Boron}, it can also form a moir\'{e} superlattice when aligned with graphene,
which leads to exceptional physical properties \cite{Yankowitz2012Emergence,Ponomarenko2013Cloning,Dean2013Hofstadters,Hunt2013Massive,Moon2014Electronic,Wang2015Evidence,RibeiroPalau2018Twistable,Chen2019Evidence,Oka2021Fractal, Brar2014Hybrid,Ni2015Plasmons,Huang2020Ultra,Zhang2021Experimental}. In its relaxed state, the moir\'{e} pattern in G/hBN exhibits a honeycomb domain structure \cite{Xue2011Scanning,Decker2011Local,Woods2014Commensurateincommensurate,SanJose2014Spontaneous,Jung2015Origin,McGilly2020Visualization},
 in contrast to a triangular pattern in homobilayers such as twisted bilayer graphene and twisted transition-metal dichalcogenides. 
 
We find that the moir\'e-phonon dispersion in G/hBN exhibits a repeating structure consisting of dispersive bands and flat bands, similarly to TBG phonon bands \cite{Koshino2019Moire}.
These characteristic structure can be described by an effective spring-mass network model, which mimics motion of the domain walls.
In particular, we show that the flat-band phonon modes are interpreted as vibrations of decoupled strings with open boundary condition,
where different flat bands correspond to different fundamental vibrating modes of a single string.
We apply the same effective model to TBG,
and we find that the flat phonon bands in TBG \cite{Koshino2019Moire}, 
correspond to decoupled strings with {\it closed} boundary condition, in contrast to the open boundary condition in G/hBN.
The difference reflects the distinction between triangular and hexagonal geometric structures of the domain wall networks.

We also calculate angular momentum of the moir\'e phonons in G/hBN. 
Generally, a phonon has a finite angular momentum in a system with broken inversion symmetry \cite{Suri2021Chiral,Maity2022Chiral}. 
We find that high amplitudes of angular momentum are concentrated near the Brillouin zone corners, where small gaps opened by the inversion symmetry breaking. We also observe significant amplitudes in the two lowest moir\'e 
phonon modes, which happen to be nearly degenerate with a small energy spacing.

This paper is organized as follows. We begin with the description of moir\'{e} superlattice in G/hBN system and introduce a continuum method to calculate long-wavelength phonons in Sec.~\ref{sec:2-methods}. 
In Sec.~\ref{sec:3-results}, we calculate phonon modes, where we reproduce the qualitative feature using the effective spring-mass model, and explain the origin of the flat phonon bands.
The phonon angular momentum is caluclated in Sec.~\ref{sec:3-angmom}.
A comparison with the TBG moir\'e phonons is given in Sec.~\ref{sec:4-tbg}. We briefly conclude the paper in Sec.~\ref{sec:5-conc}.

\section{\label{sec:2-methods}Methods}
\subsection{Geometry of moir\'{e} superlattice}
\begin{figure}
    \centering
    \includegraphics[width=\columnwidth]{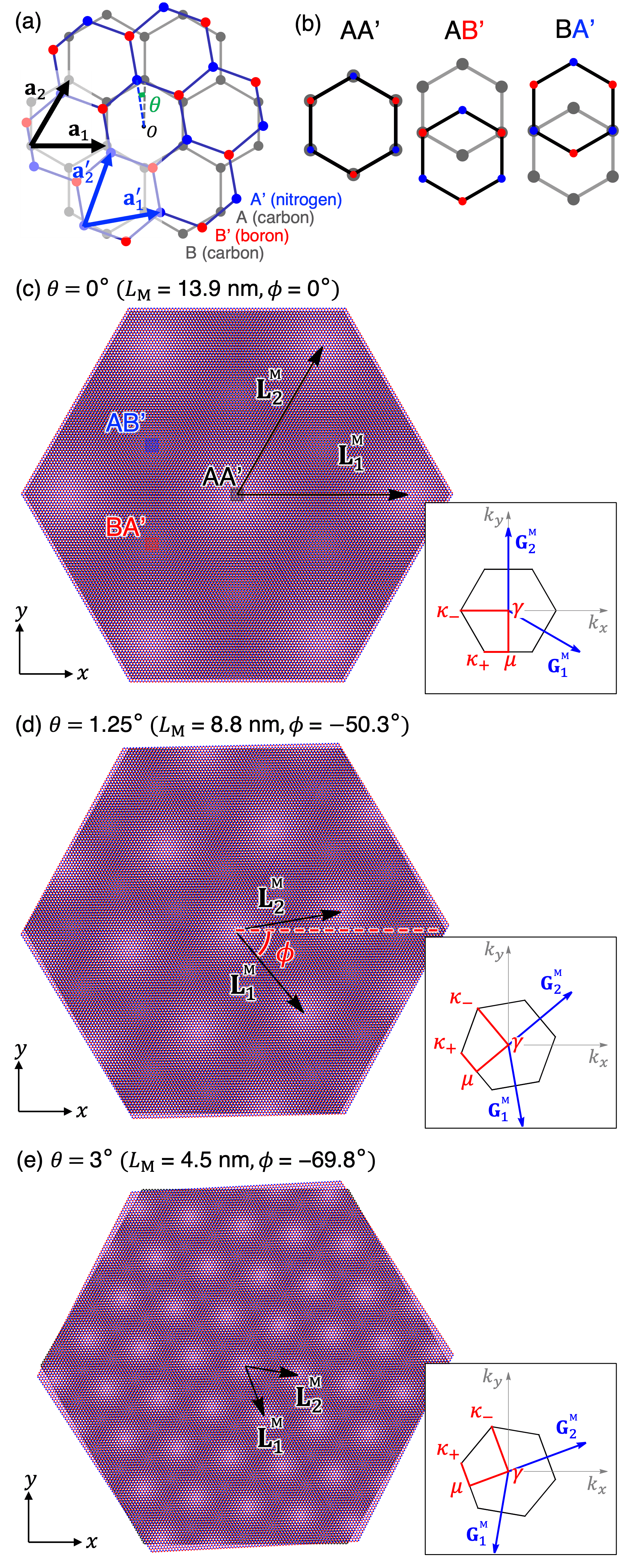}
    \caption{(a) Schematic diagram for the G/hBN system. Due to misalignments between the two layers, the local stacking structure varies between $AA'$, $AB'$, and $BA'$ illustrated in (b). These variation defines the moir\'{e} pattern as shown in (c)-(e) for $\theta=0^\circ$, $1.25^\circ$, and $3^\circ$, respectively, each with inset showing the first Brillouin zone of the superlattice.}
    \label{fig:moire-geometry}
\end{figure}

We consider a twisted bilayer system comprised of a hBN layer and a graphene layer stacked on top of each other as illustrated in Fig.~\ref{fig:moire-geometry}(a). Both graphene and hBN have two-dimensional honeycomb lattice structure, where graphene has carbon atoms at $A$ and $B$ sublattices while hBN has boron and nitrogen atoms at the $A'$ and $B'$, respectively.
The two layers have a slight lattice mismatch, approximately $\varepsilon\equiv \left(a'-a\right)/a\approx 1.8 \%$ where $a\approx0.246$ nm is graphene's lattice constant and $a'\approx0.2504$ nm is hBN's lattice constant \cite{Moon2013Optical}. 

We define an untwisted graphene-hBN bilayer ($\theta=0$) by aligning the center of a particular honeycomb cell from each layer at the origin $(x,y)=(0,0)$. The hBN layer is then rotated by an angle $\theta$ around the origin, to construct a twisted bilayer system.
The lattice constant difference and the relative twist create a moir\'{e} pattern which is periodic at a larger scale.
The primitive lattice vectors of graphene are defined as $\mathbf{a}_1= a (0,1)$ and $\mathbf{a}_2= a (1/2,\sqrt{3}/2)$,
and those of hBN are given by $\mathbf{a}'_i = \hat{M}\hat{R}\,\mathbf{a}_i \, (i=1,2)$ with isotropic expansion matrix $\hat{M}(\varepsilon)=(1+\varepsilon)\hat{I}$ and rotation matrix $\hat{R}(\theta)$. 
The reciprocal lattice vectors of graphene and hBN, denoted by 
$\mathbf{b}_i$ and $\mathbf{b}'_i$, 
respectively, satisfying $\mathbf{a}_i \cdot \mathbf{b}_j = \mathbf{a}'_i \cdot \mathbf{b}'_j = 2\pi \delta_{ij}$.

\begin{figure}
    \centering
    \includegraphics[width=\columnwidth]{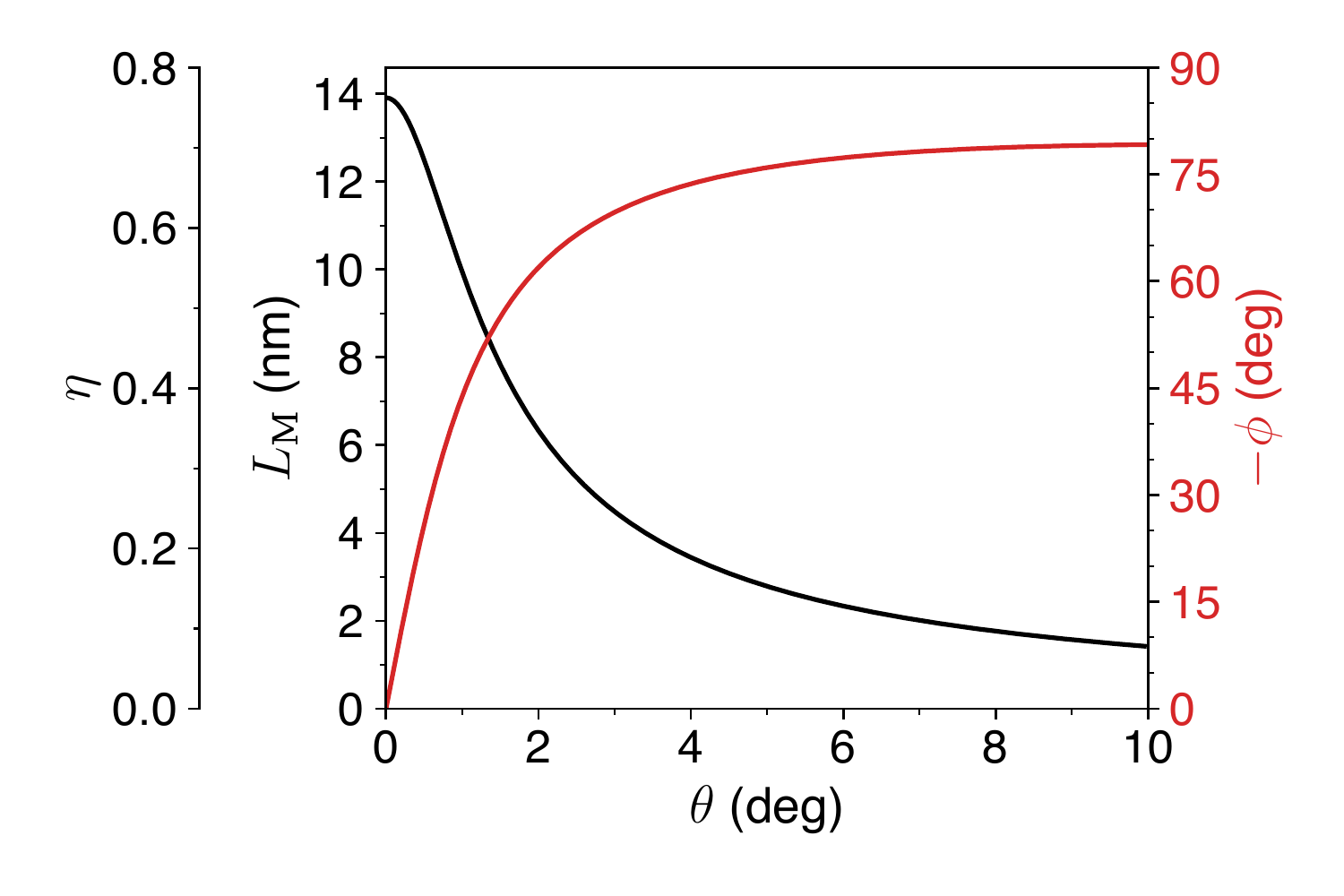}
    \caption{Variation of moir\'{e} superlattice period ($L_\text{M}$) and orientation ($\phi$) over twist angle ($\theta$). The second vertical axis on the left is the dimensionless parameter $\eta$ defined in Eq.~\ref{eq:eta}.}
    \label{fig:lm-phi}
\end{figure}

The reciprocal lattice vectors for a long-range moir\'{e} pattern is given by
$\mathbf{G}^\text{M}_i = \mathbf{b}_i - \mathbf{b}'_i$.
The corresponding real space lattice vectors $\mathbf{L}^\text{M}_i$ are obtained by the condition
$\mathbf{L}^\text{M}_i\cdot\mathbf{G}^\text{M}_j =2\pi\delta_{ij}$.
The moir\'{e} superlattice period $L_\text{M}=|\mathbf{L}^\text{M}_1|=|\mathbf{L}^\text{M}_2|$ is written as
\begin{equation}\label{eq:moire-period}
    L_\text{M} = a \frac{1+\varepsilon}{\sqrt{\varepsilon^2 + 2(1+\varepsilon)(1-\cos{\theta})}},
\end{equation}
while the angle from $\mathbf{a}_i$ to $\mathbf{L}^\text{M}_i$ defines the superlattice orientation,
\begin{equation}\label{eq:moire-angle}
    \phi = \arctan\left(
    \frac{-\sin\theta}{1+\varepsilon-\cos\theta}
    \right).
\end{equation}
Figure \ref{fig:lm-phi} shows the dependence of the superlattice period $L_\text{M}$ (black line) and orientation $\phi$ (red line) on twist angle from $0^\circ$ to $10^\circ$. 

For later convenience, we define the third reciprocal lattice vectors as 
$\mathbf{b}_3 = - \mathbf{b}_1 -\mathbf{b}_2$, 
$\mathbf{b}'_3 = - \mathbf{b}'_1 -\mathbf{b}'_2$ 
and
$\mathbf{G}^\text{M}_3=-\mathbf{G}^\text{M}_1-\mathbf{G}^\text{M}_2$,
where the three vectors of $i=1,2,3$ are trigonally symmetric. 

\subsection{Continuum methods}
\label{sec:continuum}
We describe the moir\'{e} phonons in graphene/hBN using a continuum method.
Specifically, we express the Lagrangian as a functional of smoothly-varying lattice displacement field (shifts of atoms) and obtain the Euler-Lagrange equation. 
The Lagrangian is given by $L=T-(U_E+U_B)$ with kinetic energy $T$, the elastic energy $U_E$ and the interlayer binding energy $U_B$.
In the following, we extend the formulation for TBG \cite{Nam2017Lattice, Koshino2019Moire}
to hetero moir\'e bilayers consisting of different 2D materials.

The interlayer binding energy $U_B$ is expressed as integration of the binding energy depending on the local interlayer configuration.
As a simple example, let us consider a one-dimensional system composed of two parallel atomic chains with different lattice constants.
We describe the atomic periodicities of chain 1 and 2 by sinusoidals $\cos bx$ and $\cos b'x$, respectively, where minima of the functions represent the atomic positions. We assume  $|b-b'| \ll b, b'$, i.e., the moir\'e period is much longer than the atomic periods.
The local structure at position $x$ is characterized by the phase difference between the two sinusoidals, $\varphi(x) = (b-b')x$.
Here $\varphi = 0$ represents a
perfectly overlapping arrangement 
where atoms of chain 1 and 2 are aligned,
while $\varphi=\pi$ is a staggered configuration where the atoms are aligned with the midpoint of bonds of the other chain.
The local inter-chain binding energy can be written as $V[\varphi(x)]$, a functional of the local phase difference.
The $V[\varphi]$ must be a periodic function satisfying $V[\varphi+2\pi] =V[\varphi]$.

Now we consider the lattice distortion parallel to the chain, described by smooth displacement field $u(x)$ and $u'(x)$ for chain 1 and 2, respectively.
Then the sinusoidal functions are changed to
$\cos b (x - u(x))$ and $\cos b’ (x-u’(x))$,
and hence the phase difference at $x$ becomes
\begin{align}\label{eq_phi1D}
\varphi(x ) &= b (x-u(x)) - b' (x - u'(x))
\nonumber\\
&= G^M (x - u^+(x)/2)  + \bar{b}\, u^-(x),
\end{align}
where $G^M = b-b', \bar{b} = (b+b')/2$, and $u^\pm= u'\pm u$ are interlayer-symmetric and asymmetric components of the displacement.
The local inter-chain binding energy in the presence of the distortion is given by $V[\varphi(x)]$
with $\varphi(x)$ of Eq.~\eqref{eq_phi1D}.

The binding energy between graphene and hBN can be described in a parallel manner.
The periodicity of individual honeycomb lattices are modelled by $\sum_{j=1}^3 \cos \mathbf{b}_j\cdot\mathbf{r}$ and 
$\sum_{j=1}^3 \cos \mathbf{b}'_j\cdot\mathbf{r}$
for graphene and hBN, respectively,
where minima represent atomic positions.
The local interlayer arrangement is characterized by the phase difference $(\varphi_1, \varphi_2)$, where $\varphi_j(\mathbf{r}) = (\mathbf{b}_j - \mathbf{b}'_j)\cdot\mathbf{r}$ 
for a rigid lattice without distortion.
Here $(\varphi_1, \varphi_2) = (0,0)$, $(2\pi/3,2\pi/3)$ and $(4\pi/3,4\pi/3)$ correspond to $AA'$, $AB'$ and $BA'$ stacking, respectively.
Due to 120$^\circ$ symmetry of the system, the local binding energy should be expressed 
as a symmetric function of
$\varphi_1, \varphi_2,\varphi_3(=-\varphi_1-\varphi_2)$.
In the lowest harmonics, it is written as 
\begin{equation}
    V[\varphi_1, \varphi_2] = \sum_{j=1}^3 2V_0 \cos\left[\varphi_j+\varphi_0\right] + V_{\rm const}.
\end{equation}
The parameters $V_0 = 0.202$ eV/nm$^2$, $V_{\rm const} = -0.700$ eV/nm$^2$, and $\varphi_0 = 0.956$ are obtained from the binding energies at three local alignments of $AA'$, $AB'$, $BA'$ [Fig.~\ref{fig:moire-geometry}(b)], which are 0, $-100$, and $-10$ meV per unit cell, respectively \cite{SanJose2014Spontaneous}. 

Now we consider smooth, in-plane displacement fields
$\mathbf{u}^{(1)}(\mathbf{r},t)$
and
$\mathbf{u}^{(2)}(\mathbf{r},t)$
for graphene and hBN, respectively,
which represent atomic shifts at the position $\mathbf{r}$ and time $t$.  
We also define the symmetric and antisymmetric components as
 $\mathbf{u}^\pm(\mathbf{r},t) = \mathbf{u}^{(2)}(\mathbf{r},t)\pm\mathbf{u}^{(1)}(\mathbf{r},t)$. 
The phase difference becomes
\begin{align}\label{eq:phasediff}
    \varphi_j(\mathbf{r},t) &= \mathbf{b}_j\cdot\left(\mathbf{r}-\mathbf{u}^{(1)}(\mathbf{r},t)\right) - \mathbf{b}'_j\cdot\left(\mathbf{r}-\mathbf{u}^{(2)}(\mathbf{r},t) \right) \nonumber \\ &=\mathbf{G}^\text{M}_j\cdot\left(\mathbf{r}-\mathbf{u}^+(\mathbf{r},t)/2\right)+\bar{\mathbf{b}}_j\cdot\mathbf{u}^-(\mathbf{r},t),
\end{align} 
where 
$\bar{\mathbf{b}}_j=(\mathbf{b}_j+\mathbf{b}'_j)/2$.
The total interlayer binding energy is then calculated by taking the integral over the system,
\begin{equation}
 U_B =  \int V[\varphi_1(\mathbf{r},t),\varphi_2(\mathbf{r},t)]\,d^2\mathbf{r}.
\end{equation}

The elastic energy cost associated with the in-plane distortion is described by a standard expression \cite{Suzuura2002Phonons,Nam2017Lattice},
\begin{align}\label{eq:2-uel}
    U_E = \sum_{l=1}^2 \frac{1}{2}\int
    &(\lambda^{(l)}+\mu^{(l)})\left(u^{(l)}_{xx} + u^{(l)}_{yy}\right)^2 \nonumber \\
    &+\mu^{(l)}\left[\left(u^{(l)}_{xx} - u^{(l)}_{yy}\right)^2+4\left(u^{(l)}_{xy}\right)^2\right]d^2\mathbf{r},
\end{align}
where $u_{ij}^{(l)} = (\partial_i u_j^{(l)} + \partial_j u_i^{(l)})/2$ is the strain tensor, and $\lambda^{(1)} = 3.25$ eV/\AA$^2$ and $\mu^{(1)} = 9.57$ eV/\AA$^2$ are the Lam\'{e} parameters for graphene and $\lambda^{(2)} = 3.5$ eV/\AA$^2$ and $\mu^{(2)} = 7.8$ eV/\AA$^2$ for hBN \cite{Zakharchenko2009Finite, Sachs2011Adhesion, Jung2015Origin}. Meanwhile, time-dependent displacement field gives a kinetic energy which is expressed as
\begin{equation}\label{eq:2-kinetic}
   T = \sum_{l=1}^2\int\frac{1}{2}\rho^{(l)}\left(\dot{u}_x^{(l)2}+ \dot{u}_y^{(l)2}\right)d^2\mathbf{r},
\end{equation}
where the mass density for graphene and hBN are $\rho^{(1)} = 7.61 \times 10^{-8}$ g/cm$^2$ and $\rho^{(2)} = 7.59 \times 10^{-8}$ g/cm$^2$, respectively.

The Lagrangian of the moir\'e bilayer system is given by $L = T - (U_E + U_B)$ which is a functional of the displacement vector fields $\mathbf{u}^{(l)}(\mathbf{r},t)$.
We rewrite the Lagrangian in terms of the symmetric and antisymmetric displacement vector fields $\mathbf{u}^\pm$. 
The Euler-Lagrange equation for 
$\mathbf{u}^\pm$ is obtained as
\begin{multline}\label{eq:E-L}
    \frac{1}{2}\left[\begin{pmatrix} \rho & {\rho'} \\ {\rho'} & \rho \end{pmatrix} \frac{\partial^2}{\partial t^2}
    +\begin{pmatrix} 
    \hat{K} & \hat{K}' \\
    \hat{K}' & \hat{K}
  \end{pmatrix}\right] \begin{pmatrix}\mathbf{u}^+ \\ \mathbf{u}^- \end{pmatrix}\\
    = \sum_{j=1}^3 
    2V_0 \sin\left[\varphi_j(\mathbf{r},t)+\varphi_0\right]
    \begin{pmatrix}-\mathbf{G}^\text{M}_j/2
    \\
    \bar{\mathbf{b}}_j\end{pmatrix},
\end{multline}
where 
\begin{align} \label{eq:K_matrix_derivative}
    & \hat{K} =
    -\begin{pmatrix}
    (\lambda + 2\mu) \partial_x^2 +\mu \partial_y^2 & (\lambda + \mu) \partial_x \partial_y \\
   (\lambda + \mu) \partial_x \partial_y & (\lambda + 2\mu) \partial_y^2 + \mu \partial_x^2
   \end{pmatrix},
   \\
   \label{eq:K'_matrix_derivative}
    & \hat{K}' =
    -\begin{pmatrix}
    (\lambda' + 2\mu') \partial_x^2 +\mu' \partial_y^2 & (\lambda' + \mu') \partial_x \partial_y \\
   (\lambda' + \mu') \partial_x \partial_y & (\lambda' + 2\mu') \partial_y^2 + \mu' \partial_x^2
    \end{pmatrix},
\end{align}
and
\begin{align}\label{eq:bar-tilde}
    \lambda&=\frac{\lambda^{(2)}+\lambda^{(1)}}{2}, \quad {\lambda'}=\frac{\lambda^{(2)}-\lambda^{(1)}}{2},
    \notag\\
   \mu&=\frac{\mu^{(2)}+\mu^{(1)}}{2}, \quad {\mu'}=\frac{\mu^{(2)}-\mu^{(1)}}{2}, 
   \notag\\
    \rho &=\frac{\rho^{(2)}+\rho^{(1)}}{2}, \quad {\rho'}=\frac{\rho^{(2)}-\rho^{(1)}}{2}.
\end{align}
Note that $\rho$ and $\rho'$ in Eq.~\eqref{eq:E-L}
are multiplied by a $2\times 2$ unit matrix.

We see that $\lambda', \mu', \rho'$ are responsible for the hybridization of interlayer symmetric component $\mathbf{u}^+$ and anti-symmetric component $\mathbf{u}^-$.
In our graphene-hBN system, $\lambda', \mu', \rho'$ are much smaller than $\lambda, \mu, \rho$, respectively, and hence we neglect these hybridization terms hereafter.
We also note that the effect of the moir\'e interlayer coupling [the right-hand side of Eq.~\eqref{eq:E-L}]
is much greater for $\mathbf{u}^-$  than for $\mathbf{u}^+$ since $|\bar{\mathbf{b}}_j| \gg |{\mathbf{G}}^{\rm M}_j|$ in long-range moir\'e superlattice.
Therefore, the superlattice reconstruction of the phonon bands mainly occurs for the antisymmetric modes, while it gives only a minor effect on symmetric modes.
In the following, we concentrate on the interlayer anti-symmetric modes 
$\mathbf{u}^-$.

\subsubsection{Static solution}

We assume a solution for the anti-symmetric mode in the form of \cite{Koshino2019Moire}
\begin{equation}\label{eq:upertb}
    \mathbf{u}^-(\mathbf{r},t) = \mathbf{u}^-_0(\mathbf{r}) + \delta\mathbf{u}^-(\mathbf{r},t),
\end{equation}
where 
$\mathbf{u}^-_0(\mathbf{r})$ is the static equilibrium part and $\delta\mathbf{u}^-(\mathbf{r},t)$ is a time-dependent perturbation from the equilibrium.
The equation for the static solution $\mathbf{u}^-_0(\mathbf{r})$ is given by 
setting $\delta\mathbf{u}^-(\mathbf{r},t)= 0$ in Eq.~\eqref{eq:upertb}.
Here we assume that $\mathbf{u}^-_0$ has the same periodicity as the original moir\'e pattern, and write it as
\begin{align}
& \mathbf{u}^-_0(\mathbf{r}) = 
   \sum_{\mathbf{G}} \mathbf{u}^-_{0,\mathbf{G}} e^{i\mathbf{G}\cdot\mathbf{r}},
\end{align}
where $\mathbf{G} = m\mathbf{G}^\text{M}_1+n\mathbf{G}^\text{M}_2$ are the moir\'{e} reciprocal lattice vectors.
Eq.~\eqref{eq:E-L} then becomes 
\begin{align} \label{eq:u_0}
\hat{K}_{\mathbf{G}} \mathbf{u}^-_{0,\mathbf{G}} = 
\sum_{j=1}^3 4V_0 f^j_{\mathbf{G}}
\bar{\mathbf{b}}_j,
\end{align}
where 
\begin{align} \label{eq:K_matrix}
    \hat{K}_\mathbf{q} =
    \begin{pmatrix}
    (\lambda + 2\mu) q_x^2 +\mu q_y^2 & (\lambda + \mu) q_x q_y \\
   (\lambda + \mu) q_x q_y & (\lambda + 2\mu) q_y^2 + \mu q_x^2
    \end{pmatrix},
\end{align}
and $f_\mathbf{G}^j$ is defined by
\begin{equation}\label{eq_sin_FT}
    \sin\left[\mathbf{G}^\text{M}_j\cdot\mathbf{r} + 
    \bar{\mathbf{b}}_j\cdot\mathbf{u}^{-}_0(\mathbf{r}) + \varphi_0 \right] = \sum_\mathbf{G}
    f_\mathbf{G}^j e^{i\mathbf{G}\cdot\mathbf{r}}.
\end{equation}

We solve a set of equations \eqref{eq:u_0} and \eqref{eq_sin_FT} iteratively as follows \cite{Nam2017Lattice}.
For a given $\mathbf{u}^{-}_0$, we obtain the Fourier component $f_\mathbf{G}^j$ by Eq.~\eqref{eq_sin_FT}. We then obtain the $\mathbf{u}^{-}_0$ of the next generation by $\mathbf{u}^-_{0,\mathbf{G}} = 
\sum_{j=1}^3 4V_0 f^j_{\mathbf{G}}
\hat{K}^{-1}_{\mathbf{G}} \bar{\mathbf{b}}_j$ [Eq.~\eqref{eq:u_0}].
We iterate the process until the solution converges.

\begin{figure}
    \centering
    \includegraphics[width=\columnwidth]{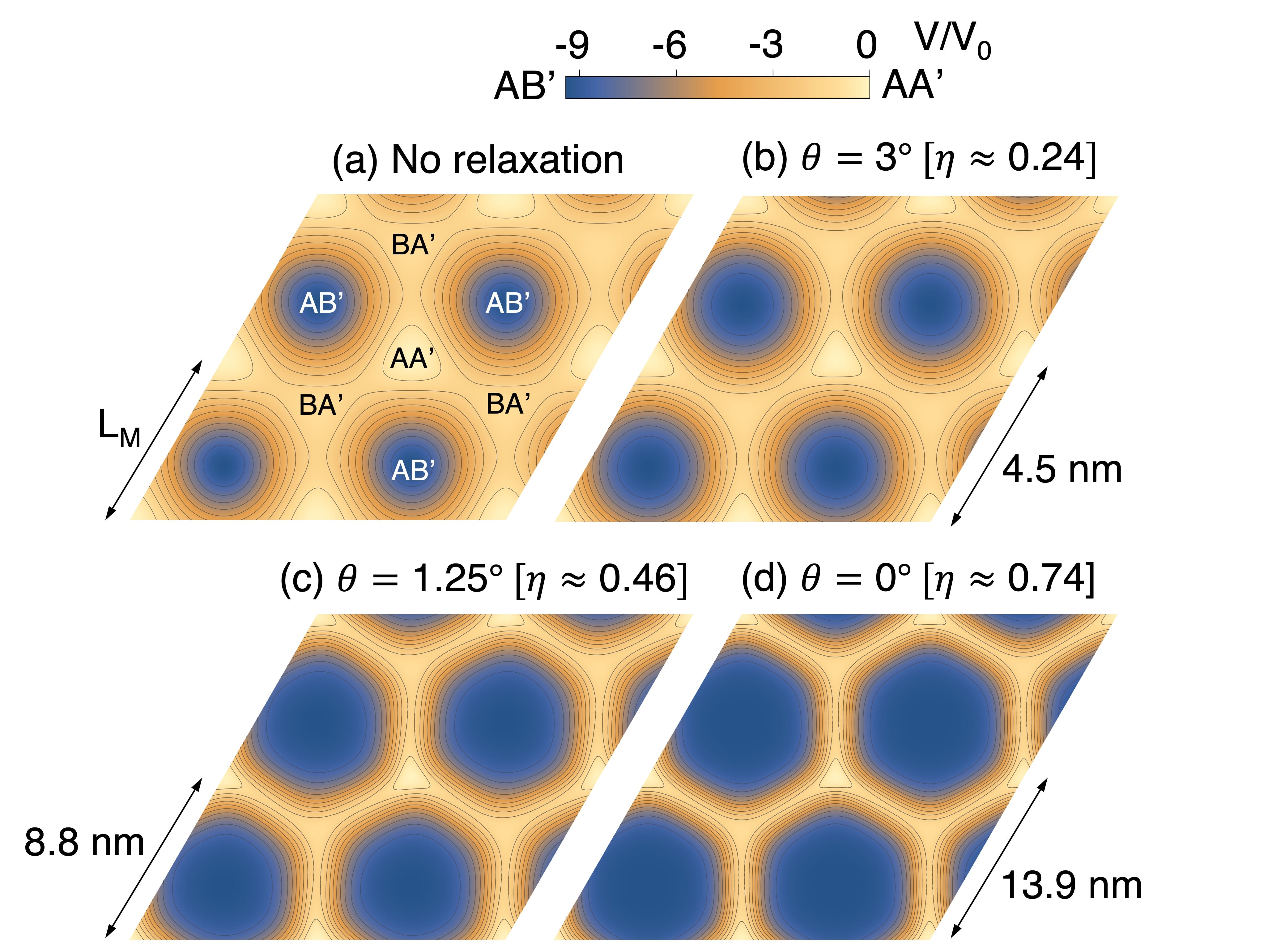}
    \caption{The equilibrium structure for (b) $\theta = 0^\circ$, (c) $1.25^\circ$, and (d) $0^\circ$ in comparison to (a) the rigid case.}
    \label{fig:static}
\end{figure}

Figure \ref{fig:static} shows the contour map of the interlayer binding energy $V[\varphi_1(\mathbf{r}),\varphi_2(\mathbf{r})]$ in the optimized  $\mathbf{u}^-_0(\mathbf{r})$ for $\theta = 3^\circ, 1.25^\circ,$ and $0^\circ$ . As the system relaxes, the most stable $AB'$ local stacking region expands to achieve minimum internal energy, and it dominates the system in small twist angles. The resulting optimal structure is an honeycomb array of domain walls which connects $AA'$ and $BA'$ stacking regions \cite{SanJose2014Spontaneous,Woods2014Commensurateincommensurate,Jung2015Origin}.

The order of relevant number of harmonics in the Fourier transformation of $\mathbf{u}_0^-$ is characterized by a dimensionless parameter \cite{Nam2017Lattice},
\begin{equation}\label{eq:eta}
    \eta = \frac{L_\text{M}}{a}\sqrt{\frac{V_0}{\lambda+\mu}}.
\end{equation}
As shown in Fig.~\ref{fig:lm-phi}
The parameter $\eta$ is a function of the twist angle $\theta$, and it monotonically increases when $\theta$ is reduced.

\subsubsection{Dynamical solution}
The time dependent part in Eq.~\eqref{eq:upertb} can be expressed in a Fourier series as
\begin{eqnarray}
    \label{eq:2-FTdu}
    \delta\mathbf{u}^-(\mathbf{r},t) =
   \frac{1}{\sqrt{S}}
    \sum_\mathbf{G}\sum_\mathbf{q}
    \delta\mathbf{u}_{\mathbf{q}+\mathbf{G}}^-(t)
    e^{i(\mathbf{q}+\mathbf{G})\cdot\mathbf{r}},
\end{eqnarray}
where $S$ is the system's total area, $\mathbf{q}$ is the phonon wave vector within MBZ.
The equation of motion, Eq.~\eqref{eq:E-L}, is then written as
\begin{equation}
\rho_r \frac{d^2}{dt^2}\delta\mathbf{u}_{\mathbf{q}+\mathbf{G}}^- =- \sum_{\mathbf{G}'}\hat{D}_\mathbf{q}(\mathbf{G},\mathbf{G}')\delta\mathbf{u}^-_{\mathbf{q}+\mathbf{G}'},
\end{equation}
 where $\rho_r=\rho/2$ is the relative mass density, 
$\hat{D}_\mathbf{q}(\mathbf{G},\mathbf{G}')=(1/2)\hat{K}_{\mathbf{q}+\mathbf{G}}\delta_{\mathbf{G},\mathbf{G}'}+\hat{V}_{\mathbf{G}'-\mathbf{G}}$
is the dynamical matrix,
and $\hat{V}$ is defined as
\begin{equation}
    \hat{V}_\mathbf{G} = 
    -2V_0 
    \sum_{j=1}^3
    h_\mathbf{G}^j
    \begin{pmatrix}
    \bar{b}_{j,x}\bar{b}_{j,x} & \bar{b}_{j,x}\bar{b}_{j,y} \\
    \bar{b}_{j,y}\bar{b}_{j,x} & \bar{b}_{j,y}\bar{b}_{j,y}
    \end{pmatrix},
\end{equation}
with 
\begin{equation}\label{eq_cos_FT}
    \cos\left[\mathbf{G}^\text{M}_j\cdot\mathbf{r} + 
    \bar{\mathbf{b}}_j\cdot\mathbf{u}^{-}_0(\mathbf{r}) + \varphi_0 \right] = \sum_\mathbf{G}
    h_\mathbf{G}^j e^{i\mathbf{G}\cdot\mathbf{r}}.
\end{equation}

At a given $\mathbf{q}$, we then obtain the phonon eigen modes by solving the following eigenvalue equation,
\begin{equation}\label{eq:eigenphonon}
    \rho_r \omega_{n,\mathbf{q}}^2 \mathbf{C}_{n,\mathbf{q}}(\mathbf{G}) =\sum_{\mathbf{G}'}\hat{D}_\mathbf{q}(\mathbf{G},\mathbf{G}')\mathbf{C}_{n,\mathbf{q}}(\mathbf{G}'),
\end{equation}
where $n$ is the mode index, $\omega_{n,\mathbf{q}}$ is the eigenfrequency, and $\mathbf{C}_{n,\mathbf{q}}(\mathbf{G})=(C^x_{n,\mathbf{q}}(\mathbf{G}),C^y_{n,\mathbf{q}}(\mathbf{G}))$ is the eigenvector normalized by $\sum_{\mathbf{G}}|\mathbf{C}_{n,\mathbf{q}}(\mathbf{G})|^2=1$.

While we neglect distortion on the out-of-plane direction throughout this work,
the real G/hBN sample is expected to be corrugated as in TBG \cite{Uchida2014Atomic,Wijk2015Relaxation,Lin2018Shear}, 
since the optimal interlayer spacing is generally registry-dependent.
Accordingly the out-of-plane phonon modes (flexural phonons) would also be subject to some superlattice effect. It is expected to be relatively minor compared to the complete restoration of in-plane phonon since out-of-plane motion does not affect the moir\'e pattern unlike in-plane interlayer sliding \cite{Koshino2019Moire}.
Also, the corrugated structure may cause some finite coupling between the in-plane modes and out-of-plane modes, but it is negligible within harmonic approximations \cite{Amorim2016Novel}.

\section{\label{sec:3-results}Moir\'{e} phonons}
\subsection{Twist angle dependence}
\begin{figure*}
    \centering
    \includegraphics[width=\textwidth]{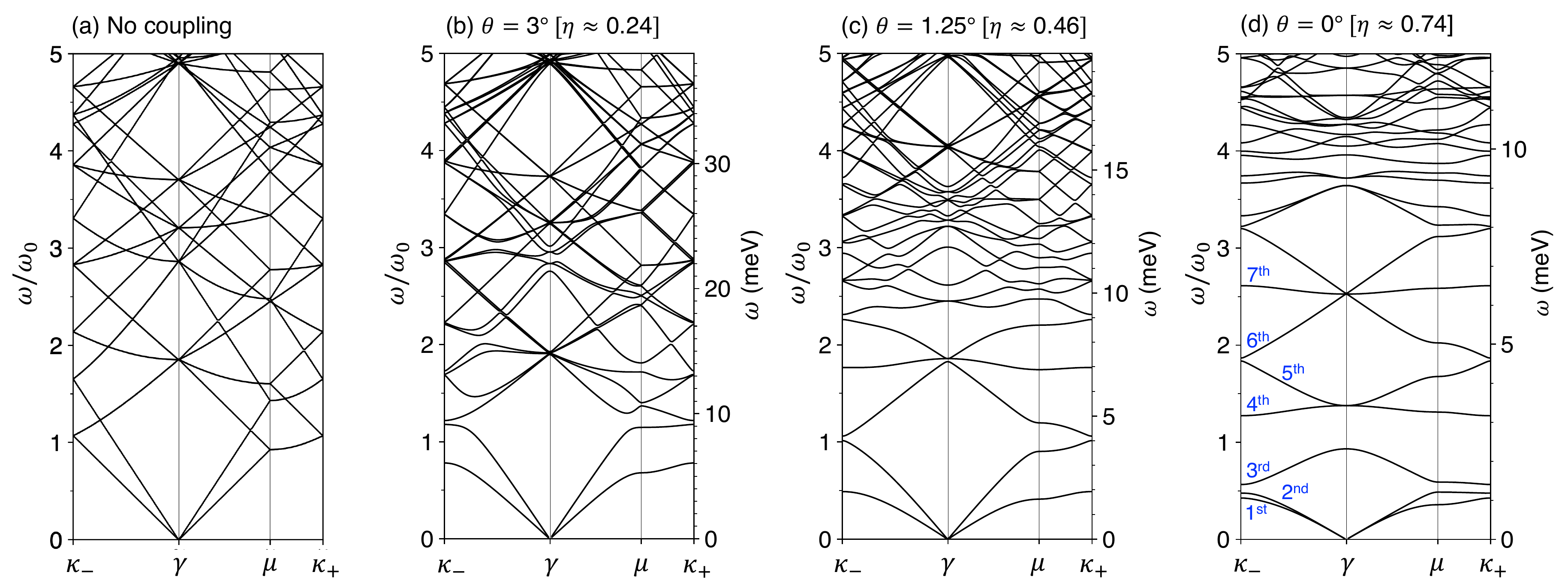}
    \caption{Phonon dispersion of the interlayer antisymmetric phonon modes for the (a) no coupling case, and three different twist angles,  (b) $\theta=3^\circ$, (c) $1.25^\circ$, and (d) $0^\circ$.}
    \label{fig:th-disp}
\end{figure*}

The calculated phonon dispersion of graphene/hBN is shown in Fig.~\ref{fig:th-disp}.
Here the panel (a) is for the case of  zero interlayer coupling, which corresponds to the empty-lattice folding of the intrinsic phonons into the MBZ. 
Figure \ref{fig:th-disp} (b)-(d) are for twist angles of $\theta = 3^\circ,1.25^\circ$ and $0^\circ$, respectively. The left vertical axis is scaled by the characteristic frequency unit,
\begin{equation}
 \omega_0= \frac{2\pi}{L_\text{M}}
 \sqrt{\frac{\lambda}{\rho}},   
\end{equation}
and the right vertical axis is in meV. The horizontal axis is scaled by $2\pi/L_\text{M}$,
where labels indicate the symmetric points of the MBZ [Fig.~\ref{fig:moire-geometry} (c-e)].
Since both the vertical and horizontal axes are scaled by $\propto 1/L_\text{M}$, we can directly compare band velocities (gradient of band lines) of different panels.
The moir\'{e} effect is observed as the appearance of gaps at the MBZ edges and the flattening of the phonon bands. As the twist angle $\theta$ is decreased below $\theta=3^\circ$, the original phonon bands are strongly modified yielding a completely different structure. At $\theta=0^\circ$, in particular, we see that the fourth are seventh bands are extremely flat in energy, which are special modes of hexagonal moir\'e systems discussed below.

\begin{figure}
    \centering
    \includegraphics[width=\columnwidth]{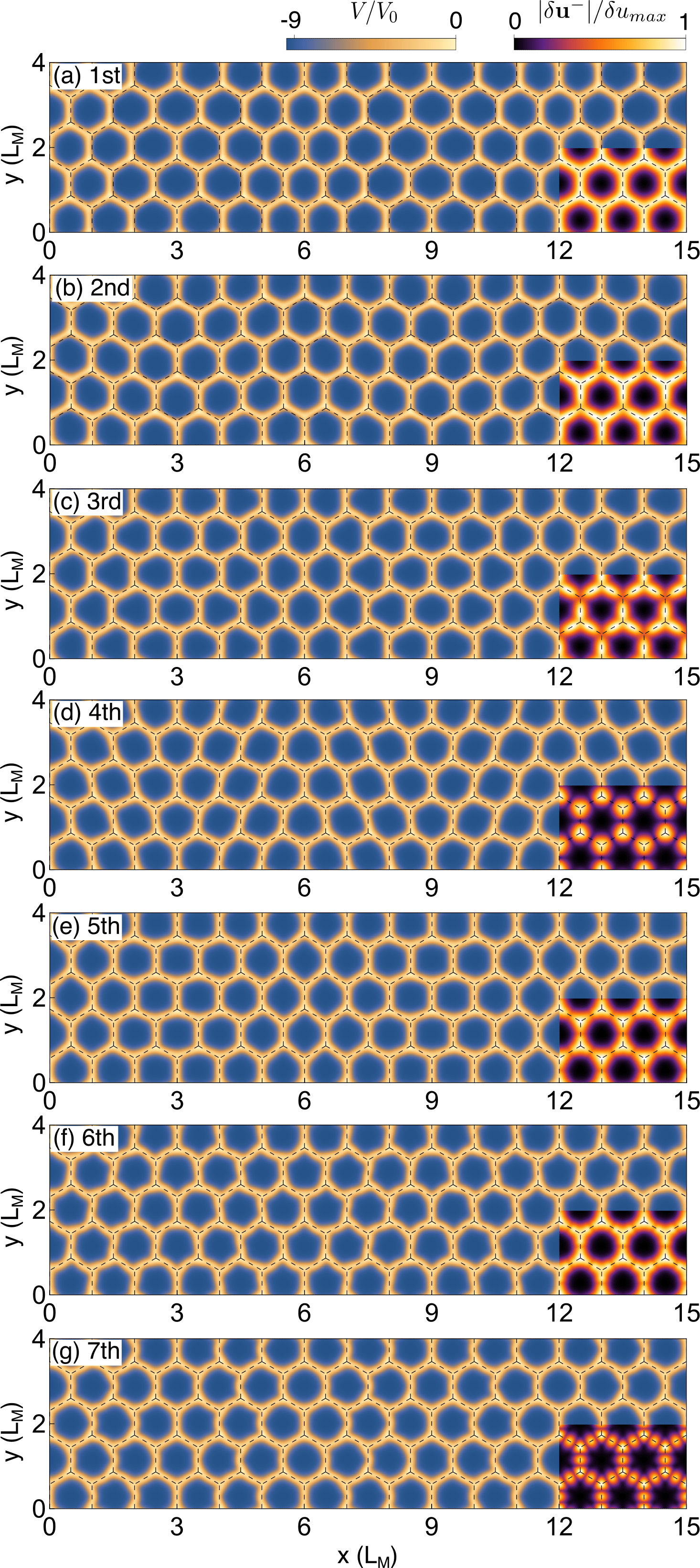}
    \caption{Phonon wave functions for the lowest 7 modes (a-h) of the interlayer antisymmetric modes in $0^\circ$ G/hBN at $\mathbf{q}=(0,\tfrac{2\pi}{6L_M})$. The color gradient represents the local binding energy. The inset at each figure shows the sum of amplitude distribution of all wave vectors within MBZ for the corresponding phonon branch.}
    \label{fig:0deg-wf}
\end{figure}

Figure \ref{fig:0deg-wf}(a-g) illustrates the phonon wave functions of the seven lowest modes of $\theta=0^\circ$ case at $\mathbf{q}=[0,2\pi/(6L_\textrm{M})]$.
The phonon modes are seen as effective oscillations at the moir\'{e} scale; for example, the lowest and the second lowest modes can be viewed as longitudinal and transverse modes of the moir\'e honeycomb lattice. 

These oscillations originate in distortion of the atomic lattice of graphene layers. In each panel of Fig.\ \ref{fig:0deg-wf}, we show the spatial distribution of the amplitude of the atomic displacement $\delta\mathbf{u}^-$ [Eq.~\ref{eq:upertb}] in the bottom right inset.
The high amplitudes are concentrated in the vicinity of the domain wall.
Importantly, we observe that the fourth and seventh modes [corresponding to the flat bands in Fig.~\ref{fig:th-disp} (d)] clearly exhibit nodes where the amplitude is completely vanishing.
As we argue in the next sections, this indicates that the wave function is composed of fundamental oscillation modes of independent strings, and it is intimately related to the band flatness. 

\subsection{Limiting case and effective model}
\label{sec:lim-eff}
\begin{figure}
    \centering
    \includegraphics[width=\columnwidth]{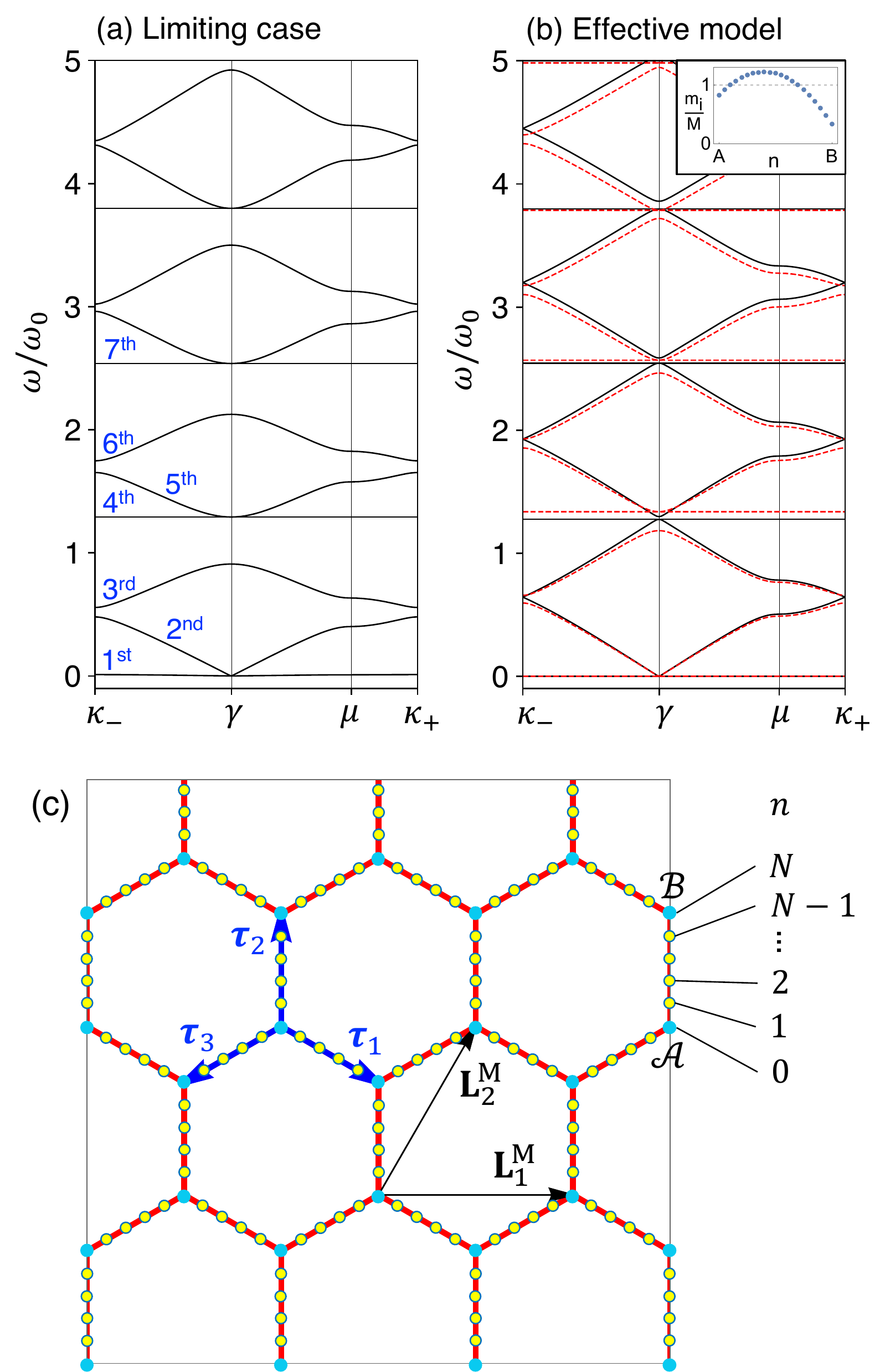}
    \caption{(a) Phonon dispersion of limiting case with $\eta\approx2$. (b) Phonon dispersion for effective model of $N=20$ with $\alpha=10$ (black line) and the inhomogeneous mass case with $\alpha=10$ (red-dashed line) which mass distribution from vertex $\mathcal{A}$ to $\mathcal{B}$ is shown in the inset. (c) Schematic diagram for the effective model. Each section of the honeycomb array is split into $N$ bonds connecting $N+1$ masses indexed from $0$ ($\mathcal{A}$) to $N$ ($\mathcal{B}$).}
    \label{fig:eff}
\end{figure}

The formation of the peculiar band structure of G/hBN moir\'e phonons can be better understood by considering a limiting case
where the parameter $\eta$ is increased with $\phi$ fixed to zero \footnote{The dependence of the dynamical equation on $\phi$ is limited to the direction of actual atomic motions, $\delta\mathbf{u}^-$, relative to the moir\'e reciprocal vectors}.
It corresponds to an imaginary situation where the interlayer binding energy $V_0$ is enhanced with the twist angle fixed to 0.
Figure \ref{fig:eff}(a) shows the phonon dispersion of the limiting case with $\eta = 2$.
We see that the same motif composed of a single completely-flat band [$(3n+1)$-th band] and two dispersive bands [$(3n+2)$-th and $(3n+3)$-th] 
appear repeatedly in the spectrum.
The fourth and seventh bands obviously correspond to the flat bands in the original $\theta=0$ model  [Fig.~\ref{fig:th-disp}].
When $\eta$ is further increased,
the low-energy dispersion (after scaling by $\omega_0$) does not change much anymore, while the periodic three-band pattern extends to higher energies. 

\begin{figure*}
    \centering
    \includegraphics[width=\textwidth]{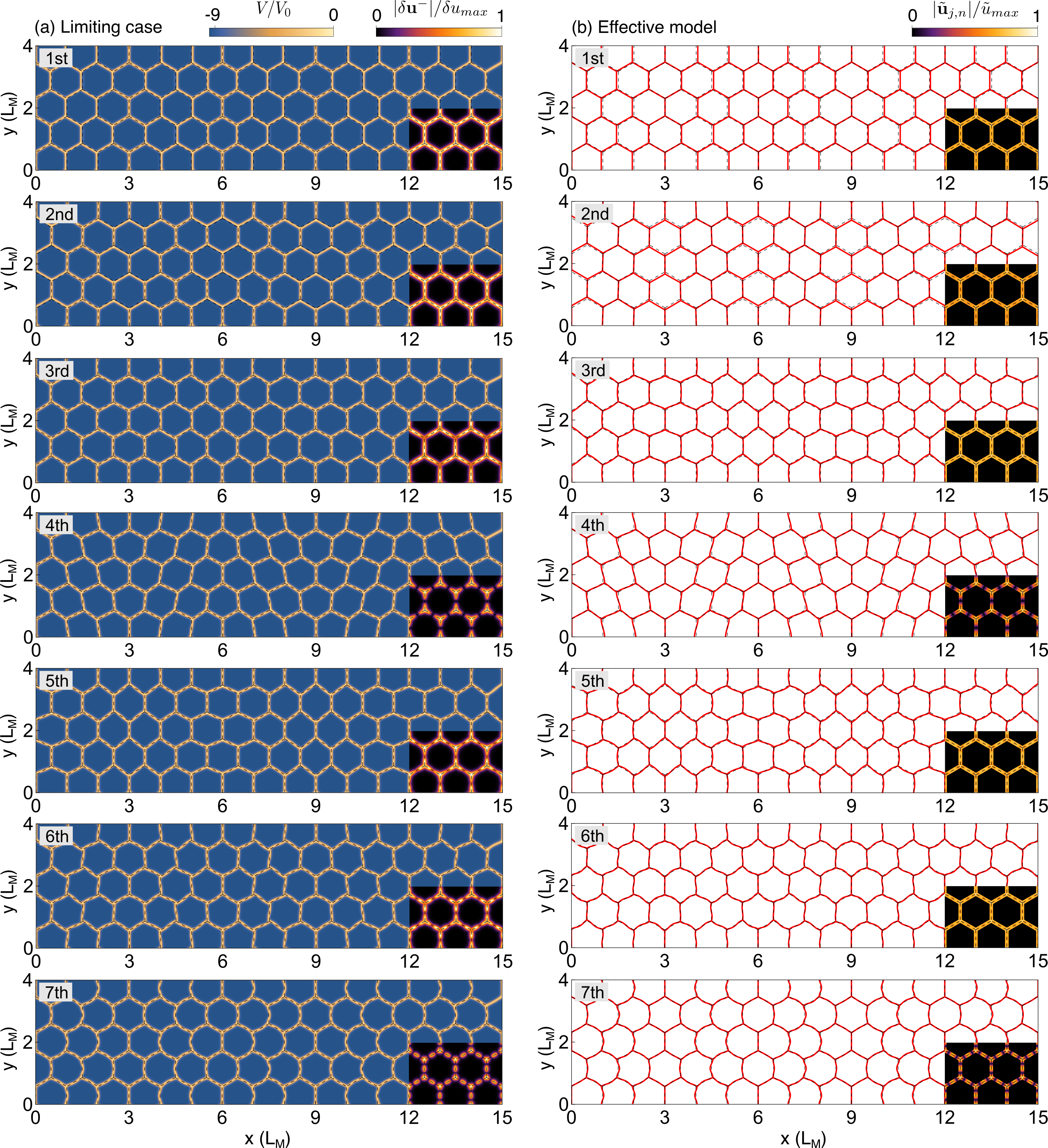}
    \caption{Phonon wave functions for the (a) limiting case and (b) effective model for the same modes and wave vectors as in Fig.~\ref{fig:0deg-wf}. The corresponding total amplitude distributions for all wave vectors in MBZ is also shown as an inset.}
    \label{fig:lim-wf}
\end{figure*}

Figure \ref{fig:lim-wf}(a) illustrates real space maps of the lowest seven modes of the limiting case $\eta=2$ at the wave number of $\mathbf{q}=[0,2\pi/(6L_\textrm{M})]$.
The patterns are basically consistent with those in Fig.~\ref{fig:0deg-wf}, while the width of the domain wall is much thinner and the system looks more like a honeycomb network of one-dimensional strings.
Since the elastic energy and the binding energy (relative to the commensurate $AB'$ stack region) are concentrated in the vicinity of the walls, the excitation energy of the phonon modes is proportional to the change of the wall length (not the squared length) relative to the static equilibrium state \cite{Koshino2019Moire,Ochoa2019Moire}. 

Based on this consideration, we construct a discrete effective model which simulates the domain wall motion with an array of masses and bonds as illustrated in Fig.~\ref{fig:eff}(c). 
A segment of the wall connecting the $AA'$ vertex (denoted as $\mathcal{A}$ sublattice) to the BA' vertex ($\mathcal{B}$ sublattice) is composed of $N$ small segments (bonds). 

The ends of each bond are linked to masses which can move on two-dimensional plane.
We define $\bm{\tau}_j$ ($j=1,2,3$) as vectors connecting $\mathcal{A}$ to the nearest $\mathcal{B}$ points [Fig.~\ref{fig:eff}(c)].
The equilibrium position of a mass is given by  
\begin{equation}
\mathbf{r}^{(j,n)}_{\mathbf{R}}=
\mathbf{R} + n \bm{\tau}_j/N,
\end{equation}
where $\mathbf{R} = m_1 \mathbf{L}^{\rm M}_1+m_2 \mathbf{L}^{\rm M}_2$ is the position of the nearest $\mathcal{A}$ sublattice, and $j=1,2,3$ represents the direction of the chain that the mass belongs to, and the index $n = 0,1,\cdots, N$ specifies the position on the chain as in Fig.~\ref{fig:eff}(c).
The displacement of the corresponding mass is denoted by ${\mathbf{u}}^{(j,n)}_{\mathbf{R}}
=(u^{(j,n)}_{x,\mathbf{R}},u^{(j,n)}_{y,\mathbf{R}})$.
This is a quantity different from the atomic displacement of the graphene lattice.
Note that three vectors $\mathbf{u}^{(j,0)}_{\mathbf{R}}(j=1,2,3)$ are actually the same variable
which represents a shift of a vertex mass at $\mathcal{A}$, and likewise $\mathbf{u}^{(j,N)}_{\mathbf{R}+\bm{\tau}_1-\bm{\tau}_j}(j=1,2,3)$  express a mass at $\mathcal{B}$.

In the presence of the displacement of masses, the change of total length of the bonds is written in the second order as
\begin{equation}
    \Delta L = \frac{1}{2l}\sum_{\mathbf{R}}\sum_{j=1}^3\sum_{n=0}^{N-1}\left[\lvert \Delta{\mathbf{u}}^{(j,n)}_{\mathbf{R}} \rvert^2 - \left(\hat{\bm{\tau}}_j\cdot\Delta
    {\mathbf{u}}^{(j,n)}_{\mathbf{R}}\right)^2 \right],
\end{equation}
where 
$\Delta{\mathbf{u}}^{(j,n)}_{\mathbf{R}}
=
{\mathbf{u}}^{(j,n+1)}_{\mathbf{R}}
-
{\mathbf{u}}^{(j,n)}_{\mathbf{R}}$
and $\hat{\bm{\tau}}_j = \bm{\tau}_j/|\bm{\tau}_j|$ is a unit vector along $j$ direction. 
Here the length change linear to ${\mathbf{u}}^{(j,n)}_{\mathbf{R}}$ is considered to be zero, assuming that an overall expansion of the whole system is restricted by the boundary condition.
The change in the total energy is then given by ${U}=\alpha V_0 w_d \Delta L$, where $w_d=(a/4)\sqrt{(\lambda+\mu)/V_0}$ is the width of the wall and $\alpha$ is a numerical constant to match the energy scale of original system \cite{Nam2017Lattice,Koshino2019Moire}. By the Fourier transform ${\mathbf{u}}^{(j,n)}_{\mathbf{R}}=\sum_\mathbf{q}
{\mathbf{u}}^{(j,n)}_\mathbf{q}
\exp(i\mathbf{q}\cdot
\mathbf{r}^{(j,n)}_{\mathbf{R}})$, 
${U}$ can be written as
\begin{equation}\label{eq:Ueff}
{U}=
\frac{K}{2}
\sum_\mathbf{q}\sum_{j=1}^3\sum_{n,n'=0}^{N}
\left[{\mathbf{u}}_{-\mathbf{q}}^{(j,n')}\right]^T\hat{D}_\mathbf{q}^j(n',n){\mathbf{u}}_\mathbf{q}^{(j,n)},
\end{equation}
where $K = \alpha V_0 w_d/l$ is the effective spring constant.
The $\hat{D}^j_\mathbf{q}(n,n')$ is a $2\times 2$ dynamical matrix of which non-zero elements are given by,
\begin{align}
&\hat{D}^j_\mathbf{q}(n,n)
=
\left\{
\begin{array}{ll}
\hat{T}_j & (n = 0,N) \\
2\hat{T}_j & (n = 1,2,\cdots,N-1),
\end{array}
\right.
\\
&\hat{D}^j_\mathbf{q}(n-1,n)=\left[\hat{D}^j_\mathbf{q}(n,n-1)\right]^\dagger
=
-\hat{T}_je^{i\mathbf{q}\cdot\tau_j/N},
\end{align}
where
\begin{align}
\hat{T}_j = 
\begin{pmatrix}
1-(\hat{\tau}^x_j)^2 & -\hat{\tau}^x_j\hat{\tau}^y_j
\\
-\hat{\tau}^x_j\hat{\tau}^y_j &
1-(\hat{\tau}^y_j)^2
\end{pmatrix}.
\end{align}
The $\hat{T}_j$ is a projection operator perpendicular to $\hat{\bm{\tau}}_j$,
which works for an arbitrary vector $\bm{x}$ as
$\hat{T}_j \bm{x} = (\delta^{\mu\nu} - \hat{\tau}^\mu_j\hat{\tau}^\nu_j)x^{\nu}= 
\bm{x} - \hat{\bm{\tau}}_j (\hat{\bm{\tau}}_j\cdot \bm{x})$.

The kinetic energy of the system is 
\begin{equation}
{T}=\frac{M}{2}\sum_\mathbf{q}
\Biggl[
\lvert\dot{{\mathbf{u}}}_\mathbf{q}^\mathcal{A}\rvert^2 
+ 
\lvert\dot{{\mathbf{u}}}_\mathbf{q}^\mathcal{B}\rvert^2 
+ 
\sum_{j=1}^3\sum_{n=1}^{N-1}\lvert\dot{{\mathbf{u}}}_\mathbf{q}^{(j,n)}\rvert^2
\Biggr],
\end{equation}
where $M=\rho a^2l/w_d$ is the effective mass\cite{Koshino2019Moire}, and
${\mathbf{u}}_\mathbf{q}^\mathcal{A} = {\mathbf{u}}_\mathbf{q}^{(j,0)}$ and ${\mathbf{u}}_\mathbf{q}^\mathcal{B} = {\mathbf{u}}_\mathbf{q}^{(j,N)}$. 
The Euler-Lagrange equation is then given by $M\ddot{\mathbf{u}}_\mathbf{q}^{(j,n)}=K\sum_{n'}\hat{D}_\mathbf{q}^j(n',n)\mathbf{u}_\mathbf{q}^{(j,n')}$, which is solved to obtain eigen phonon frequencies and the corresponding wave functions.

The phonon dispersion of the effective model with $N=20$ and $\alpha=10$ is shown as the black line in Fig.~\ref{fig:eff}(b). 
We see that the effective model qualitatively reproduces the flat bands and the three-band periodic pattern observed in the limiting case [Fig.~\ref{fig:eff}(a)].
We also see a perfect correspondence 
of the wave functions
between the limiting case [Fig.~\ref{fig:lim-wf}(a)] and the effective model Fig.~\ref{fig:lim-wf}(b)].

On the other hand, the effective model does not capture the gap opening at $\gamma$ and $\kappa$, which is observed in the original model
[Fig.~\ref{fig:eff}(a)].
This can be qualitatively incorporated by introducing inhomogeneous mass distribution in a single chain.
For example, we assume
\begin{equation}
    M_n/M = 1 - b_1\left(n/N - b_2\right)^2,
\end{equation} 
where $b_1$ and $b_2$ are tunable parameters. The resulting phonon dispersion with $(b_1,b_2)=(2,0.4)$ is shown in Fig.~\ref{fig:eff}(b) as red-dashed line, 
with the mass distribution is shown in the inset. 
Here the gap opening at $\kappa$ point results from the inversion symmetry breaking 
by the asymmetric mass distribution.
In the original G/hBN system, the inversion symmetry is broken by the inequivalent binding potential on the $AA'$ and $BA'$ stacking regions, which creates sub-meV gap at the $\kappa$ points [Fig.~\ref{fig:th-disp}]. 
The broken inversion symmetry also gives rise to chiral moir\'{e} phonons \cite{Suri2021Chiral,Maity2022Chiral} which will be discussed further in Sec.~\ref{sec:3-angmom}.
The gap opening at the $\gamma$ point is not related to the inversion symmetry, but it is caused by difference in matching of the mass distribution and the wave-amplitude distribution.

\subsection{Origin of the flat bands}

To consider the origin of the flat bands,
we take the $N=1$ case and obtain the analytical solution. Here a unit cell contains masses only at $\mathcal{A}$ and $\mathcal{B}$,  and hence the equation has only four degrees of freedom. 
The equation of motion is written as
\begin{align}\label{eq:eigen_eq}
M \omega^2
\begin{pmatrix}
 {{\mathbf{u}}}_\mathbf{q}^\mathcal{A}\\
 {{\mathbf{u}}}_\mathbf{q}^\mathcal{B}
\end{pmatrix}
=
K\sum_{j=1}^3
\begin{pmatrix}
\hat{T}_j & -\hat{T}_je^{i\mathbf{q}\cdot\tau_j} \\
 -\hat{T}_je^{i\mathbf{q}\cdot\tau_j} & \hat{T}_j
\end{pmatrix}
 \begin{pmatrix}
 {{\mathbf{u}}}_\mathbf{q}^\mathcal{A}\\
 {{\mathbf{u}}}_\mathbf{q}^\mathcal{B}
\end{pmatrix},
\end{align}
where $\omega$ is the eigen frequency.
The obtained phonon dispersion has a similar structure to the lowest four bands of $N=20$ model, where flat bands appear in
the first and fourth bands with
eigen frequencies $\omega=0, 3\sqrt{K/M}$, respectively.

The corresponding eigenstates are given by
\begin{equation}\label{eq:flat-eigstates}
\begin{pmatrix}
{\mathbf{u}}_\mathbf{q}^\mathcal{A} \\ {\mathbf{u}}_\mathbf{q}^\mathcal{B}
\end{pmatrix}
= \begin{pmatrix}
\mathbf{f}_\mathbf{q} \\ \mp\mathbf{f}^*_\mathbf{q}
\end{pmatrix},
\end{equation}
respectively, where 
\begin{equation}
\mathbf{f}_\mathbf{q} = \sum_{j=1}^3 \bm{\tau}_j
e^{-i\mathbf{q}\cdot\bm{\tau}_j}.
\end{equation}
It is straightforward to check that Eq.~\eqref{eq:flat-eigstates} satisfies the eigen equation Eq.~\eqref{eq:eigen_eq}, by using the relation
\begin{align}\label{eq_tau_project}
\hat{T}_j \mathbf{f}_\mathbf{q}
=
-\hat{T}_j \mathbf{f}^*_\mathbf{q} e^{i\mathbf{q}\cdot\bm{\tau}_j},
\end{align}
and $\sum_{j=1}^3 \hat{T}_j = (3/2)I$, where $I$ is a 2$\times$2 unit matrix.

The expression of Eq.~\eqref{eq:flat-eigstates} leads to an important observation for the motion of the neighboring masses.
Let us consider a pair of masses at $\mathcal{A}$ and $\mathcal{B}$ points separated by $\bm{\tau}_j$.
According to Eq.~\eqref{eq:flat-eigstates},
the motions of the two points are given by
${\mathbf{u}}^\mathcal{A}(\mathbf{R}) = C \mathbf{f}_\mathbf{q}$ and
${\mathbf{u}}^\mathcal{B}(\mathbf{R}+\bm{\tau}_j) = \mp C \mathbf{f}_\mathbf{q}^* e^{i\mathbf{q}\cdot\bm{\tau}_j}$,
where $C$ is a common constant.
Using Eq.~\eqref{eq_tau_project}, we immediately have
\begin{equation}
\hat{T}_j {\mathbf{u}}^\mathcal{A}(\mathbf{R})
=
\pm\hat{T}_j {\mathbf{u}}^\mathcal{B}(\mathbf{R}+\bm{\tau}_j),
\end{equation}
for the first and fourth modes, respectively.
Noting that $\hat{T}_j$ is the projection operator perpendicular to $\bm{\tau}_j$,
we conclude that, in the flat band modes,
the neighboring vertices $\mathcal{A}$ and $\mathcal{B}$
always move either in phase (the first mode) or out of phase (the fourth mode)
when the motion is projected perpendicularly to the bond. 

\begin{figure}
    \centering
    \includegraphics[width=0.7\columnwidth]{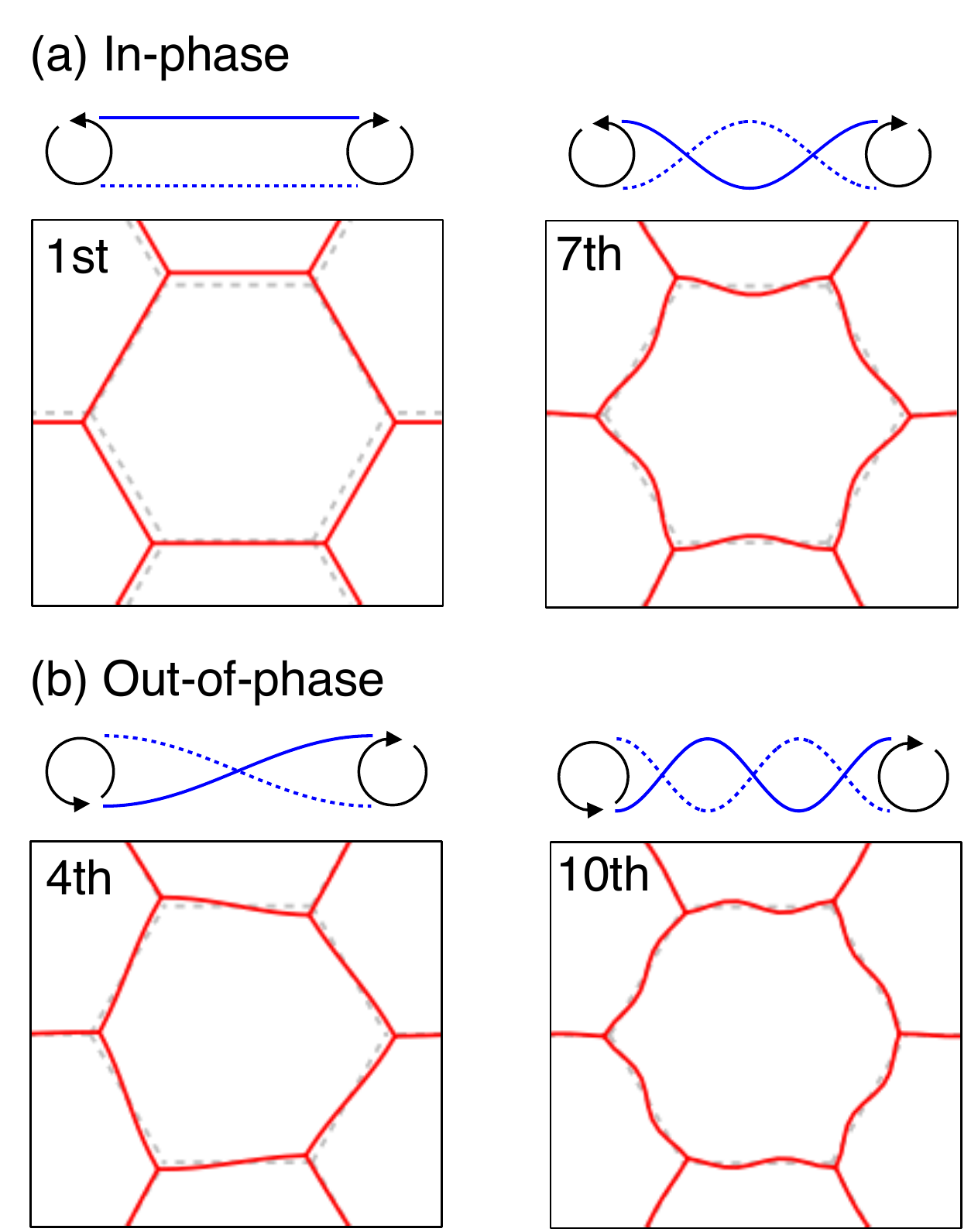}
    \caption{Phase synchronization for the perpendicular motion of neighboring vertices in the effective model: (a) in-phase mode, and (b) out-of-phase mode.}
    \label{fig:phase-synchro}
\end{figure}

Actually, this relationship holds for vertex-site motions of any flat band modes in $N\geq 1$ cases,
where $6n+1$-th and $6n+4$-th modes are associated with the in-phase and out-of-phase motions, respectively,
as illustrated in Fig.~\ref{fig:phase-synchro}.
The phase synchronization of the vertex sites means that masses in each single chain can collectively vibrate as a stationary wave of an isolated string.
Since the phase synchronization persists at any $\mathbf{q}$ as shown above, this gives a flat dispersion at the frequency of the corresponding fundamental mode of the string.
The vertex motions parallel to bonds are not generally synchronized, but they are irrelevant for the band flatness because the contributions of the parallel shifts to the total bond length cancel as a whole, and do not change the total energy.
Here note that the energy of an effective spring is linearly proportional to its length as argued above.

A notable feature in moir\'e phonons in the limiting model (and the corresponding effective model) is the existence of a flat band at zero frequency.
The complete flattening of the lowest 
branch implies that the regular honeycomb array is unstable against expansion/contractions of the hexagonal unit cell.
This can be understood by noting that we can modify a regular honeycomb array into an irregular pattern
without a change in the total length of the domain wall (and hence the total energy), by expanding/shrinking hexagons with the orientation of sides (domain walls) kept unchanged.
The situation is quite similar to solid phases of adsorbed atoms on a graphite surface,
where the commensurate domain wall formation was discussed \cite{Villain1980Commensurate}.
In the real G/hBN superlattice, the lowest band is not completely flat as seen in Fig.~\ref{fig:th-disp}(d), and therefore the regular honeycomb superlattice is energetically stable. The finite dispersion of the lowest band would be incorporated by adding vertex-vertex interaction energy in the effective model.

\section{Angular momentum}
\label{sec:3-angmom}

In a system without inversion symmetry, the phonons generally acquire a chiral nature with finite angular momentum and the Berry curvature \cite{Zhang2015Chiral,Suri2021Chiral,Maity2022Chiral}.
In the current system, 
the inversion symmetry breaking term enters as
a small difference in local binding energy for $BA'$ and $AB'$ stacking structures [Fig.~\ref{fig:static}],
which is caused by the inversion-asymmetric structure of hBN. To clarify the existence of chiral phonons in G/hBN, we calculate the out-of-plane component of angular momentum defined as \cite{Zhang2014Angular,Suri2021Chiral}
\begin{equation}\label{eq:Lz-real}
     L^z = \rho \int d^2\mathbf{r} \sum_{l=1}^{2}\left(\delta\mathbf{u}^{(l)} \times \delta\dot{\mathbf{u}}^{(l)}\right)_z,
\end{equation}
where $l(=1,2)$ is the layer index and
$\delta\mathbf{u}^{(l)}(\mathbf{r},t)$ is the displacement vector of layer $l=1,2$.

By using the Fourier transformation of the displacement vector [Eq.~\ref{eq:2-FTdu}] and the relation $\delta\mathbf{u}^{(2)}=-\delta\mathbf{u}^{(1)}=(1/2)\delta\mathbf{u}^{-}$, we can rewrite Eq.~\eqref{eq:Lz-real} as,
\begin{equation}
    \label{eq:Lz-recip}
    L^z = \sum_\mathbf{q}\sum_\mathbf{G}\left(\delta\mathbf{u}_{\mathbf{q}+\mathbf{G}}^-\times\delta\mathbf{p}^-_{\mathbf{q}+\mathbf{G}}\right)_z,
\end{equation}
where $\delta\mathbf{p}_{\mathbf{q}}^-
=\rho_r\delta\dot{\mathbf{u}}^-_{-\mathbf{q}}$.
In terms of phonon creation and annihilation operators, $\delta\mathbf{u}_{\mathbf{q+G}}^-$ and $\delta\mathbf{p}_{\mathbf{q+G}}^-$ are written as \cite{Koshino2020Effective},
\begin{align}
  &  \delta\mathbf{u}_{\mathbf{q+G}}^-=\sum_n
    \mathbf{C}_{n,\mathbf{q}}(\mathbf{G})
    \sqrt{ \frac{\hbar}{2\rho_r\omega_{n,\mathbf{q}}} }
    (a_{n,\mathbf{q}} + a^\dagger_{n,-\mathbf{q}}),
\nonumber\\
  &  \delta\mathbf{p}_{\mathbf{q+G}}^- = \sum_n
    i\mathbf{C}^*_{n,\mathbf{q}}(\mathbf{G})
    \sqrt{ \frac{\hbar\rho_r\omega_{n,\mathbf{q}}}{2} }
    (a^\dagger_{n,\mathbf{q}} - a_{n,-\mathbf{q}}),
\end{align}
where $\mathbf{C}_{n,\mathbf{q}}(\mathbf{G})$ is the normalized eigenvector of Eq.~\eqref{eq:eigenphonon}.
Substituting these into Eq.~\ref{eq:Lz-recip}, we have
\begin{multline}
    L^z = \frac{i\hbar}{2}
    \sum_{\mathbf{q},\mathbf{G}}\sum_{n,n^\prime}
    \sqrt{ \frac{\omega_{n^\prime,\mathbf{q}}}{\omega_{n,\mathbf{q}}} }
    \left[\mathbf{C}_{n,\mathbf{q}}(\mathbf{G}) \times \mathbf{C}^*_{n^\prime,\mathbf{q}}(\mathbf{G})\right]_z
    \\
    \times(a_{n,\mathbf{q}} + a^\dagger_{n,-\mathbf{q}})
    (a^\dagger_{n,\mathbf{q}} - a_{n,-\mathbf{q}}).
\end{multline}
Finally, the expectation value in equilibrium is written as
\begin{align}
    &\label{eq:Lz-2ndQ}
    \langle L^z \rangle = \sum_{n,\mathbf{q}}
    L^z_{n,\mathbf{q}}
    \left[ f(\omega_{n,\mathbf{q}}) + \frac{1}{2} \right], 
\end{align}
where
\begin{align}
    &L^z_{n,\mathbf{q}} = i\hbar \sum_\mathbf{G}\mathbf{C}_{n,\mathbf{q}}(\mathbf{G})\times\mathbf{C}^*_{n,\mathbf{q}}(\mathbf{G}),
\end{align}
and $f(\omega) = 1/(\exp(\hbar\omega/k_BT) - 1)$ is the Bose-Einstein distribution function,
and we note that
$\langle a^\dagger_{n,\mathbf{q}} a_{n^\prime,\mathbf{q}^\prime} \rangle = f(\omega_{n,\mathbf{q}})\delta_{n,n^\prime}\delta_{\mathbf{q},\mathbf{q}^\prime}$,
$\langle a_{n,\mathbf{q}} a_{n^\prime,\mathbf{q}^\prime} \rangle =
\langle a^\dagger_{n,\mathbf{q}} a^\dagger_{n^\prime,\mathbf{q}^{'
}} \rangle = 0$, and $\omega_{n,\mathbf{q}}=\omega_{n,-\mathbf{q}}$.

\begin{figure}
    \centering
    \includegraphics[width=0.9\columnwidth]{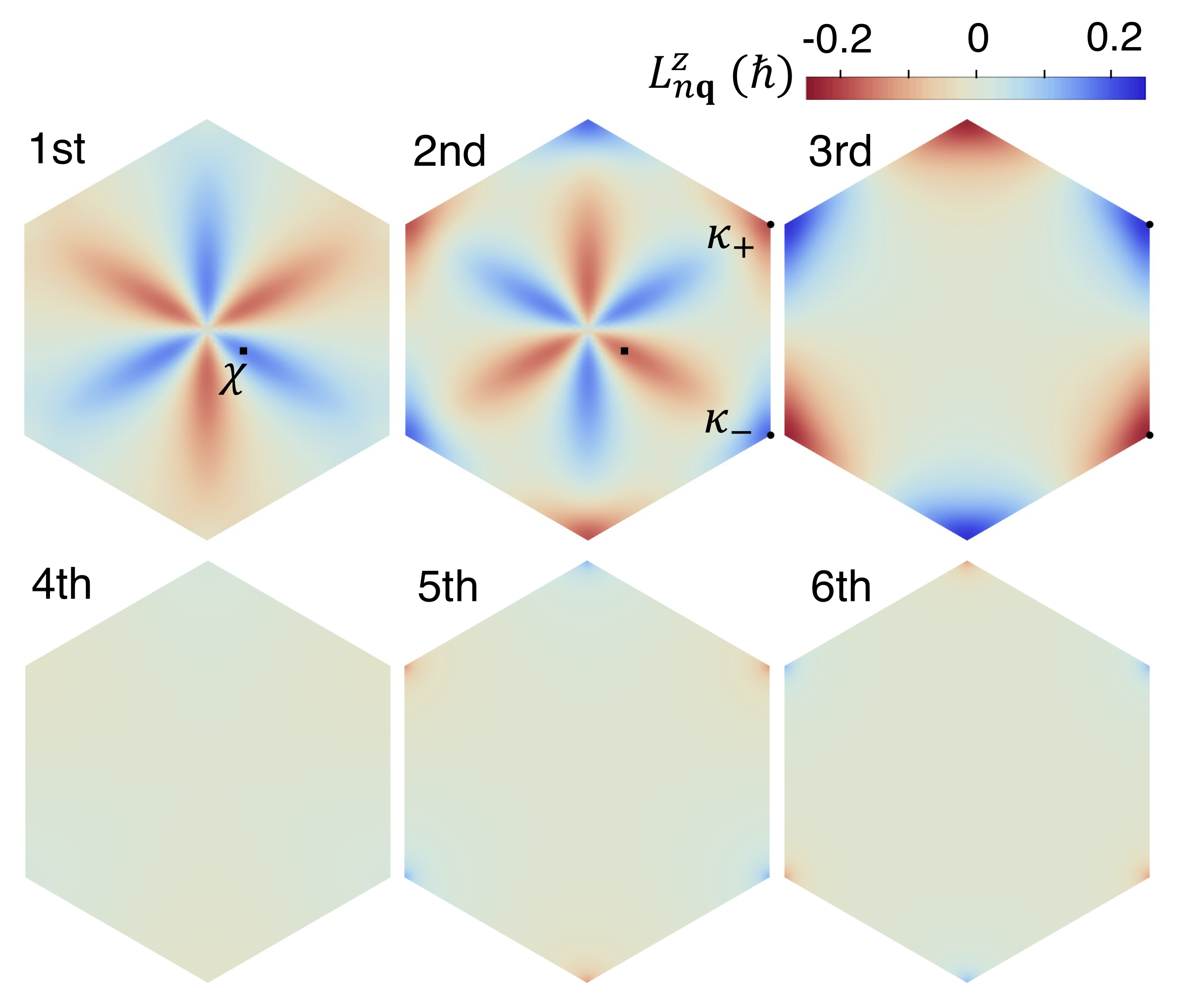}
    \caption{Angular momentum for the lowest sixth bands of $0^\circ$ G/hBN within the MBZ.}
    \label{fig:angmom1}
\end{figure}

\begin{figure}
    \centering
    \includegraphics[width=0.9\columnwidth]{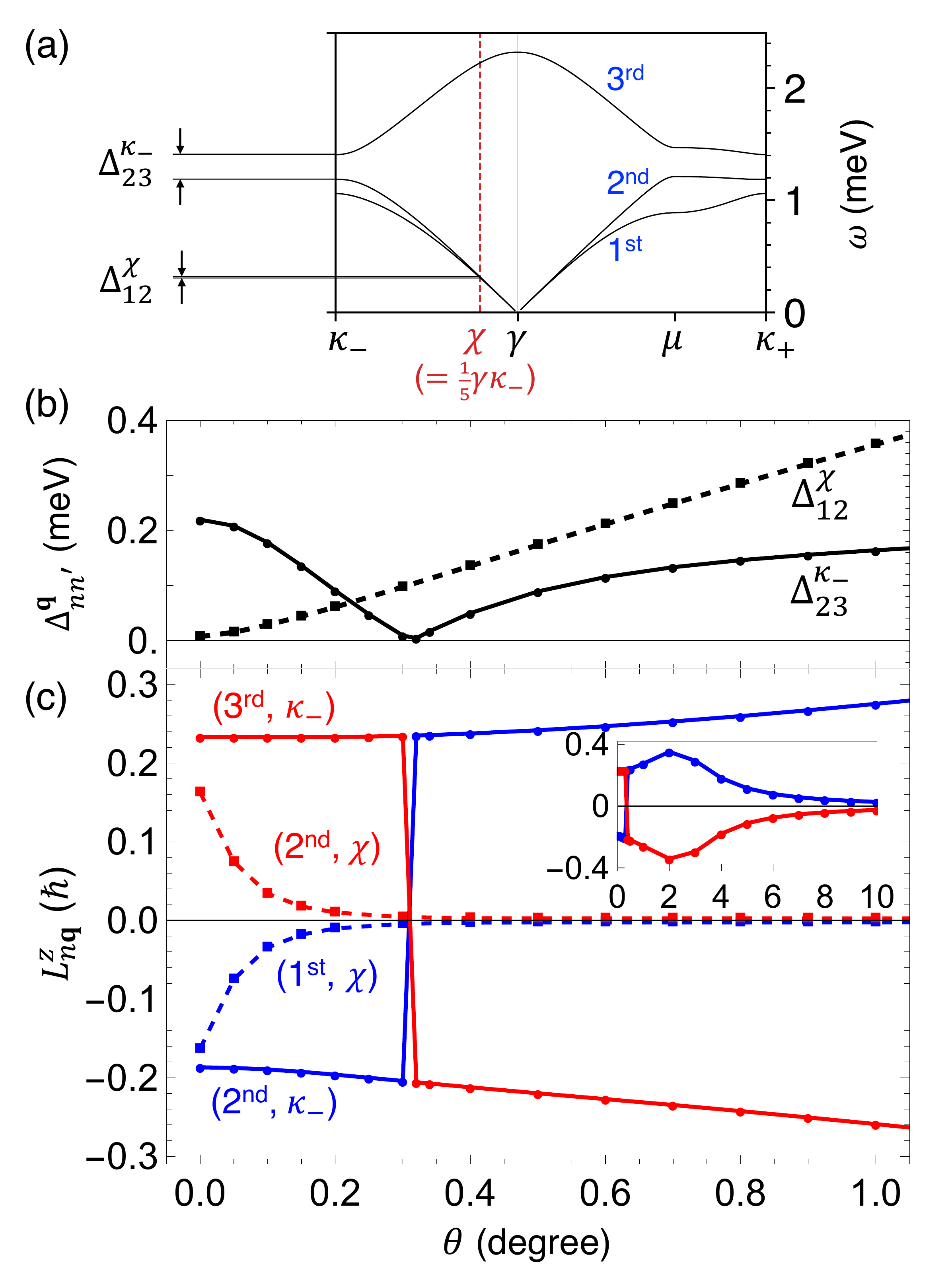}
    \caption{(a) Dispersion of the lowest three bands of $0^\circ$ G/hBN. (b) Twist angle dependence of gap width ($\Delta_{nn'}^\mathbf{q}$) between the $n$-th and $n'$-th band at $\mathbf{q}$. $\chi$ is taken as $\tfrac{1}{5}\gamma\kappa_-$. (c) Twist angle dependence of the angular momenta for each corresponding bands involved in (b) with inset showing larger range of angle up to $\theta=10^\circ$.}
    \label{fig:angmom2}
\end{figure}

Figure \ref{fig:angmom1} shows the $k$-space distribution of the angular momentum $L^z_{n,\mathbf{q}}$ for the lowest six bands in $0^\circ$-stack of G/hBN. 
We observe relatively large amplitudes with opposite signs in the second and third bands around the BZ corner $\kappa_\pm$. This corresponds to a gap opening caused by the inversion symmetry breaking [$\Delta_{23}^{\kappa_-}$, Fig.~\ref{fig:angmom2}(a)].
In the fourth to sixth bands,
notable angular momentum 
is observed only in the close vicinity of $\kappa_\pm$, in accordance with very small symmetry-breaking gaps in the phonon band structure.

Figure \ref{fig:angmom2}(c) shows the twist-angle dependence of the angular momentum
$L^z_{n,\mathbf{q}}$ of the second and the third bands at $\kappa_-$.
The corresponding plot for the gap width  $\Delta_{23}^{\kappa_-}$ is shown in  Fig.~\ref{fig:angmom2}(b). 
We observe that the angular momenta of these two bands are swapped when the gap closes at $\theta\sim 0.3^\circ$.
The absolute values peak at $\sim$2$^\circ$ and monotonically decrease in larger twist angles, as shown in the inset of
Fig.~\ref{fig:angmom2}(c).

In Fig.~\ref{fig:angmom1}, we also observe
notable signals of angular  momentum in the two lowest bands around lines connecting $\gamma$ and the $\kappa_\pm$ points. This can be attributed to a tiny energy distance
between the two bands, where perturbative matrix elements of the symmetry breaking terms give rise to sizable angular momentum by hybridizing these nearly-degenerate bands.
We present the twist angle dependence of
the angular momentum of the first and the second bands at $\chi \equiv (1/5)\gamma\kappa_-$ 
in Fig.~\ref{fig:angmom2}(c),
and also the corresponding plot of the energy distance $\Delta_{12}^{\chi}$ between the two bands at $\chi$ in Fig.~\ref{fig:angmom2}(b).
In increasing the twist angle from 0,
the $\Delta_{12}^{\chi}$ become rapidly increases, and their angular momenta immediately vanish correspondingly.
 Note that this property is not seen in TBG, where the ratio between longitudinal and transverse phonon velocity converges to $\sqrt{3}$ in the small-angle limit \cite{Koshino2019Moire,Gao2022Symmetry}.

\section{Flat phonon bands in twisted bilayer graphene}
\label{sec:4-tbg}

\begin{figure*}
    \centering
    \includegraphics[width=0.6\textwidth]{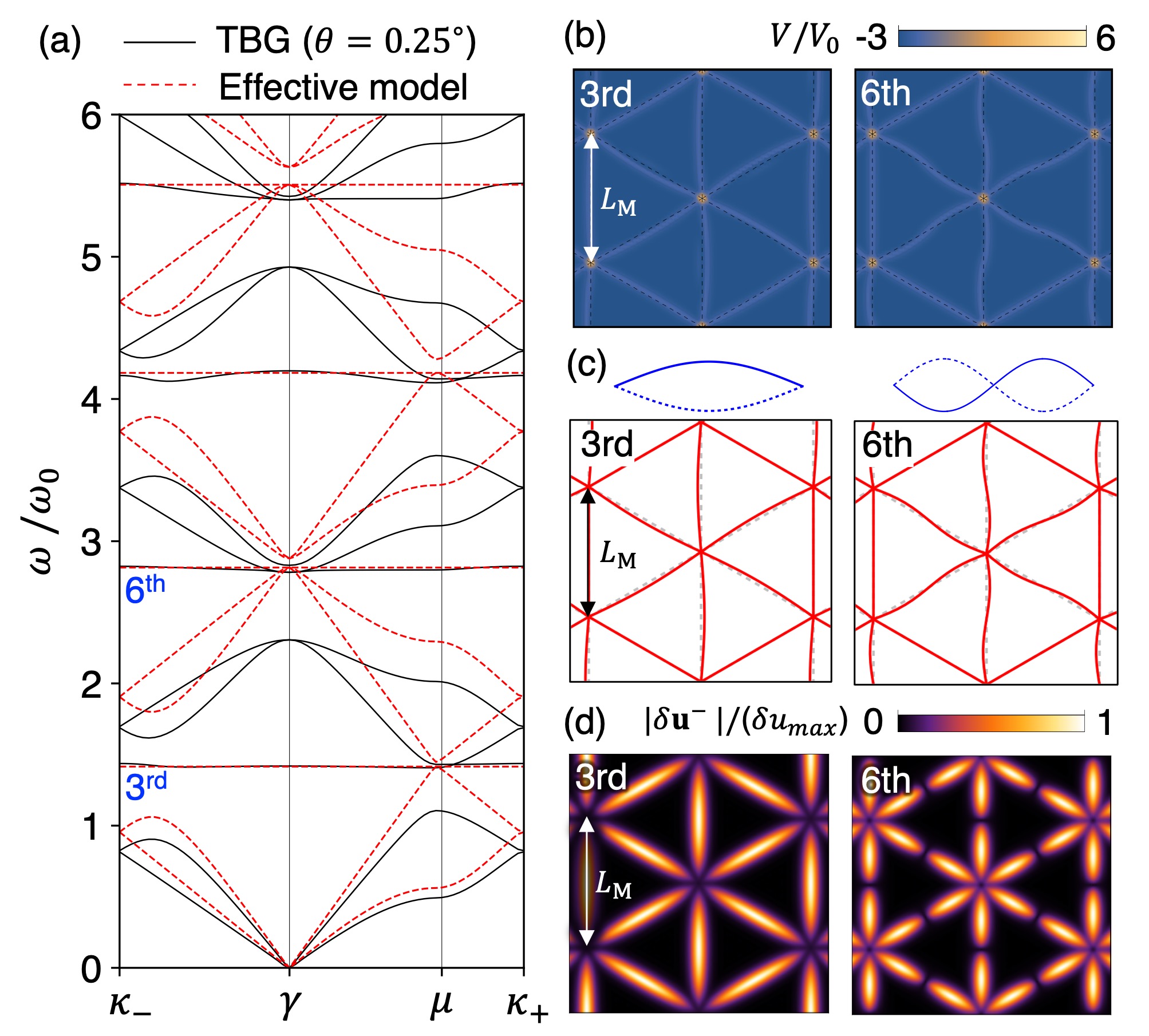}
    \caption{(a) Phonon dispersion of TBG with $\theta=0.25^\circ$ (black line) and effective triangular model with $N=15$ (red-dashed line) with corresponding wave function for the two lowest flat bands at $\kappa_-$ are given in (b) and (c) respectively. (d) The oscillation amplitudes distribution for the two lowest flat bands of the TBG.}
    \label{fig:tbg}
\end{figure*}

Moir\'{e} phonons have been previously studied for twisted bilayer graphene (TBG) by one of the authors and a co-worker \cite{Koshino2019Moire}, where it was shown that flat phonon bands emerge in small twist angles in a similar way to G/hBN system.
Here we show that the flat bands of TBG can also be understood as fundamental vibration modes of a single string, but with a different boundary condition.
In Fig.\ \ref{fig:tbg}(a), the black line represent the phonon dispersion of TBG at twist angle of $\theta = 0.25^\circ$,
which is calculated by the the same continuum model in Sec.~\ref{sec:continuum} with $\rho^{(l)}$, $\lambda^{(l)}$ and $\mu^{(l)}$ set to graphene's parameters.
Here we chose very small twist angle to achieve the limiting case with a large $\eta$.
We also construct an effective spring-mass model in a similar way to G/hBN's, except that the masses and bonds are arranged in a triangular network to be consistent with the domain walls in TBG system. The corresponding phonon dispersion for the effective model with $N=15$ and $\alpha=12$ is shown in red-dashed lines in \ref{fig:tbg}(a).
We observe that the two models have similar band structures, where flat bands appear in $3n$-th bands.

The phonon wave functions for the two lowest flat bands (3rd and 6th) in the $0.25^\circ$ TBG and the corresponding effective model are shown in Fig.~\ref{fig:tbg}(b) and (c) respectively, while the amplitudes ($\delta \mathbf{u}^-$) for the former are given in Fig.~\ref{fig:tbg}(d).
Again, the vibration of a single wall segment can be viewed as a fundamental oscillation mode of an isolated string, but now we see that the vertices ($AA'$ region) are stationary at any wavelengths. This is in contrast to the flat bands in G/hBN case, where the vertices always correspond to antinodes. The reason for this is due to the absence of sublattice in the triangular lattice, which forbids any phase synchronization for motions of vertices in a finite wave-vector $\mathbf{q}$. Therefore, stationary waves can exist only when the vertices are fixed in their position. 
To summarize, the flat bands in the G/hBN superlattice (honeycomb lattice) correspond to string-like oscillations with the open boundary condition, while ones in TBG (triangular lattice) correspond to those with the closed boundary condition.

Another notable difference 
is the zero-energy flat band which we observed in the G/hBN limiting model does not exist in TBG.
This is because, unlike a honeycomb lattice, it is impossible to distort triangular lattice without changing its total side length.

\section{Conclusion}
\label{sec:5-conc}

We have studied the characteristics of moir\'{e} phonons in G/hBN systems, particularly focusing on the origin of phonon band flattening.
By using a continuum approach, we 
demonstrate that the phonon band structure exhibits a regular pattern of flat bands and dispersive bands.
The emergence of the flat phonon bands can be reproduced by simulating the domain walls with a honeycomb array of strings, of which energy is proportional to the length. 
The flat band mode corresponds to a fundamental vibration of a single string with open boundary condition,
where the projected motions of the neighboring vertices always synchronize independently of wave vectors.
The flat phonon bands of TBG can also be understood by a similar string model with a triangular network,
where the flat phonon modes are associated with single-string vibrations with closed boundary condition due the lack of sublattice
in the triangular lattice.
These results suggest that the emergence of flat phonon bands is a general feature of the long-period moir\'{e} superlattice.

We have also calculated the phonon angular momentum. 
Our results reveal the existence of chiral phonons not only near the highly symmetric MBZ corners, but also in the entire $k$-space region for the lowest bands. While the former remains finite at large twist angle, the latter is closely related to the nearly-identical 
phonon velocities which only occurs at $\theta \sim 0^\circ$.

The flat bands in the phonon spectrum are expected to entail various physical consequences.
For instance, non-propagating phonons in the low-energy spectrum should be manifested in a considerable suppression of thermal conductivity relative to intrinsic graphene \cite{Qian2021Phonon}. 
Meanwhile, band flatness is generally associated with the existence of a spatially-localized eigenmode.
In our moir\'e phonon system, this suggests that highly localized phonon excitation (vibration of a single domain wall sector) is possible, as was achieved in photonic lattice \cite{Mukherjee2015Observation,Vicencio2015Observation}. 
Another possibility is bosonic condensation into a flat band by an external excitation, which was realized in an exciton-polariton system \cite{Baboux2016Bosonic}. 
For moir\'e phonons, a possible excitation mechanism is through electromagnetic radiation. 
Since G/hBN moir\'e super lattice has inversion-asymmetric charge densities \cite{Correa2016Ab}, the moir\'e phonon modes at the zone boundary would couple to an in-plane AC electric field.
We also expect an extension of similar calculations to other heterobilayer moir\'e systems straightforward, which could unveil broader roles of moir\'e phonons in thermal and electronic transport phenomena.

\begin{acknowledgments}
This work was supported in part by JSPS KAKENHI Grant Number JP21H05236, JP21H05232, JP20H01840, JP20H00127, and JP20K14415 and by JST CREST Grant Number JPMJCR20T3, Japan. L.~P.~A.~K.~ acknowledges support from The Konosuke Matsushita Memorial Foundation Scholarship and JST SPRING, Grant Number JPMJSP2138.
\end{acknowledgments}

\bibliography{bibliography}

\end{document}